\def\HPpage{
\documentclass[a4paper,12pt]{article}
\addtolength{\topmargin}{-2.5 true cm}
\addtolength{\textheight}{3.5 true cm}
\addtolength{\textwidth}{4 true cm} \addtolength{\hoffset}{-1
true cm} \setlength{\oddsidemargin}{0cm}
\setlength{\evensidemargin}{0cm}
}

\def\Izvpage{
\documentclass[a4paper,12pt]{article}
\addtolength{\topmargin}{-1.5 true cm}
\addtolength{\textheight}{1.5 true cm}
\addtolength{\textwidth}{1 true cm} }

\Izvpage

\usepackage{amsmath, amsthm, amsfonts, amssymb, graphicx}
\usepackage[mathscr]{eucal}

\theoremstyle{plain}
\newtheorem{theorem}{Theorem}[section]

\newtheorem{lemma}{Lemma}[section]

\theoremstyle{remark}

\numberwithin{equation}{section}

\newcommand{\sect}[1]
{\addtocounter{section}{1}

\medskip

\begin{center}
{\textbf{\large\arabic{section}. #1}}
\end{center}
\setcounter{equation}{0}\setcounter{theorem}{0}\setcounter{lemma}{0}
\setcounter{remark}{0} \smallskip }

\newcommand{\sectn}[1]
{ 
\medskip
\begin{center}
{\large\textbf{#1}}
\end{center}
\setcounter{equation}{0}\setcounter{theorem}{0}\setcounter{lemma}{0}
\setcounter{remark}{0} \smallskip }

\def\th{\theta}
\def\Om{\Omega}
\def\om{\omega}
\def\e{\varepsilon}
\def\g{\gamma}
\def\G{\Gamma}
\def\l{\lambda}
\def\p{\partial}
\def\D{\Delta}
\def\bs{\backslash}
\def\k{\varkappa}

\def\a{\alpha}

\def\t{\widetilde}
\def\h{\widehat}

\def\Si{\Sigma}
\def\d{\delta}

\newcommand{\PF}[1]
{\noindent\textbf{#1}}

\begin{document}

\allowdisplaybreaks

\def\thefootnote{}

\begin{center}
{\large\textbf{ASYMPTOTICS AND ESTIMATES OF DEGREES OF
CONVERGENCE IN THREE-DIMENSIONAL BOUNDARY VALUE PROBLEM WITH
FREQUENT INTERCHANGE OF BOUNDARY CONDITIONS }}
\end{center}

\begin{center}
{\large Denis I. Borisov}\footnotetext[1]{The work was partially
supported by RFBR (Nos. 02-01-00693, 00-15-96038) and Program
''Universities of Russia'' of Ministry of Education of Russia
(UR.04.01.010).}
\end{center}

\begin{quote}
{\small {\em  Bashkir State Pedagogical University, October
Revolution St.,~3a, 450000, Ufa, Russia. E-mail:}
\texttt{BorisovDI@ic.bashedu.ru, BorisovDI@bspu.ru}}
\end{quote}

\medskip

\begin{abstract}
We consider a singular perturbed eigenvalue problem for Laplace
operator in a cylinder with frequent interchange of type of
boundary condition on a lateral surface. These boundary
conditions are prescribed by partition of lateral surface in a
great number of narrow strips on those the Dirichlet and Neumann
conditions are imposed by turns. We study the case of the
homogenized problem containing Dirichlet condition on the
lateral surface. When the width of strips varies slowly, we
construct the leading terms of eigenelements' asymptotics
expansions. We also estimate the degree of convergence for
eigenvalues if the strips' width varies rapidly.
\end{abstract}

\sectn{Introduction}

The present paper is devoted to the studying of a
three-dimensional boundary value problem  with frequent
interchange of boundary condition. The main feature of
formulation of such problems is partition of domain's boundary
in two parts, on the first the boundary condition of one type is
imposed (ex. Dirichlet condition) while on the second the
boundary condition of another type is prescribed (ex. Neumann
condition). One of this parts is assumed to depend on a small
parameter and consist of disjoint components; moreover, the
small parameter going to zero, the number of components
increases unboundedly while the measure of each component tends
to zero. The question of homogenization for the problems of such
kind are investigated well enough (see, for instance,
\cite{Dl}-\cite{Dv}). The main homogenization result established
in the papers cited can be formulated as follows. The solutions
to the boundary value problem with frequent interchange of
boundary conditions converges to ones of the problems with
classic boundary conditions whose type is determined by a
relationship between measured of parts of boundary with
different boundary condition in the origin problem. The authors
of  \cite{Ch}, \cite{GCh}, \cite{ChOl}, \cite{Dr} considered the
interchange between Dirichlet and Neumann or Robin condition and
obtained the estimates of degrees of convergence provided each
connected component with boundary condition of one of the types
shrinks to a point. The asymptotics for the solutions of the
problems with frequent interchange were constructed in
\cite{AA}-\cite{CR}. Two-dimensional case was studied in
\cite{AA}-\cite{DAN}. In papers \cite{BDU}, \cite{BMs} they
constructed complete asymptotics expansions of Laplace
operator's  eigenelements in a circular cylinder with frequent
interchange between Dirichlet and Neumann condition imposed on
narrow strips in a lateral surface; their width was  constant.
In \cite{BDU} the author considered the case of the homogenized
problem with Dirichlet condition on a lateral surface under
additional assumption that the width of strips with Dirichlet
and Neumann condition are of same order of smallness. In
\cite{BMs} they studied the case corresponding to the
homogenized problem with Neumann or Robin condition on a lateral
surface. In both cases it was shown that original perturbed
problem has simple and double eigenvalues only. In addition, in
\cite{BMs} for cylinder of arbitrary cross-section and the width
of strips varying slowly in the case of homogenized problem with
Neumann or Robin condition on a lateral surface the author
constructed  the leading terms of asymptotics expansions for
eigenelements, where eigenvalues were supposed to converge to
simple limiting eigenvalues.

In the present paper we consider a singular perturbed eigenvalue
problem for Laplacian in a cylinder of arbitrary cross-section.
On the upper basis we impose Dirichlet condition while on the
lower one we prescribe Neumann condition. The lateral surface is
partitioned in a great number of narrow strips with varying
width governed by two character parameters. On these strips the
Dirichlet and Neumann conditions are imposed by turns. We study
the case of homogenized problem with Dirichlet condition on the
lateral surface. Provided the strips' width varies slowly we
construct the leading terms of the two-parametrical asymptotics
expansions for the eigenelements. The form of these expansions
allows us to maintain that in a general case the complete
splitting of limiting multiply eigenvalues takes place and the
perturbed problem has simple eigenvalues only. We also study the
particular case of circular cylinder and show that depending on
the strips' width both the previous situation of the complete
splitting of multiply eigenvalues and the situation of
non-splitting may arise. We adduce the sufficient condition
guaranteeing that the perturbed problem has at least one double
eigenvalue. For the case of the strips' width varying rapidly we
estimate the degree of convergence for perturbed eigenvalues.

The result of this work were announced in  \cite{CR}.

\sect{Description of the problem and formulation of the results}

Let $x'=(x_1,x_2)$, $x=(x', x_3)$ be Cartesian coordinates in
$\mathbb{R}^2$ and $\mathbb{R}^3$, $\om\subset\mathbb{R}^2$ be a
bounded simply connected domain whose boundary is infinitely
differentiable, $\Om=\om\times [0,H]$, $H>0$, $\om_1$, $\om_2$
be upper and lower basis of the cylinder $\Om$, $\om_1=\{x:
x'\in\om, x_3=H\}$, $\om_2=\{x: x'\in\om, x_3=0\}$. By $s$ we
denote the natural parameter of the curve $\p\om$. We  suppose
that  $N$ is a natural number, tending to infinity; $\e=H/(\pi
N)$ is a small parameter. We define a set $\g_\e$ located in a
lateral surface  $\Si$ of the cylinder $\Om$ and consisting of
$N$ narrow strips:
\begin{equation*}
\g_\e=\left\{x: x'\in\p\om, \left|x_3-
\e\pi(j+1/2)\right|<\e\eta\mathsf{g}_\e(s),
j=0,\ldots,N-1\right\},
\end{equation*}
where $\eta=\eta(\e)$, $0<\eta(\e)<\pi/2$, $\mathsf{g}_\e\in
C^\infty(\p\om)$ is an arbitrary function obeying an estimate
$0<c\le \mathsf{g}_\e(s)\le 1$ with constant $c$ independent on
$\e$ and $s$ (cf. fig.)

\begin{figure}[tb]
\begin{center}
\noindent
\includegraphics[width=8.84 true cm, height=9.29 true cm]{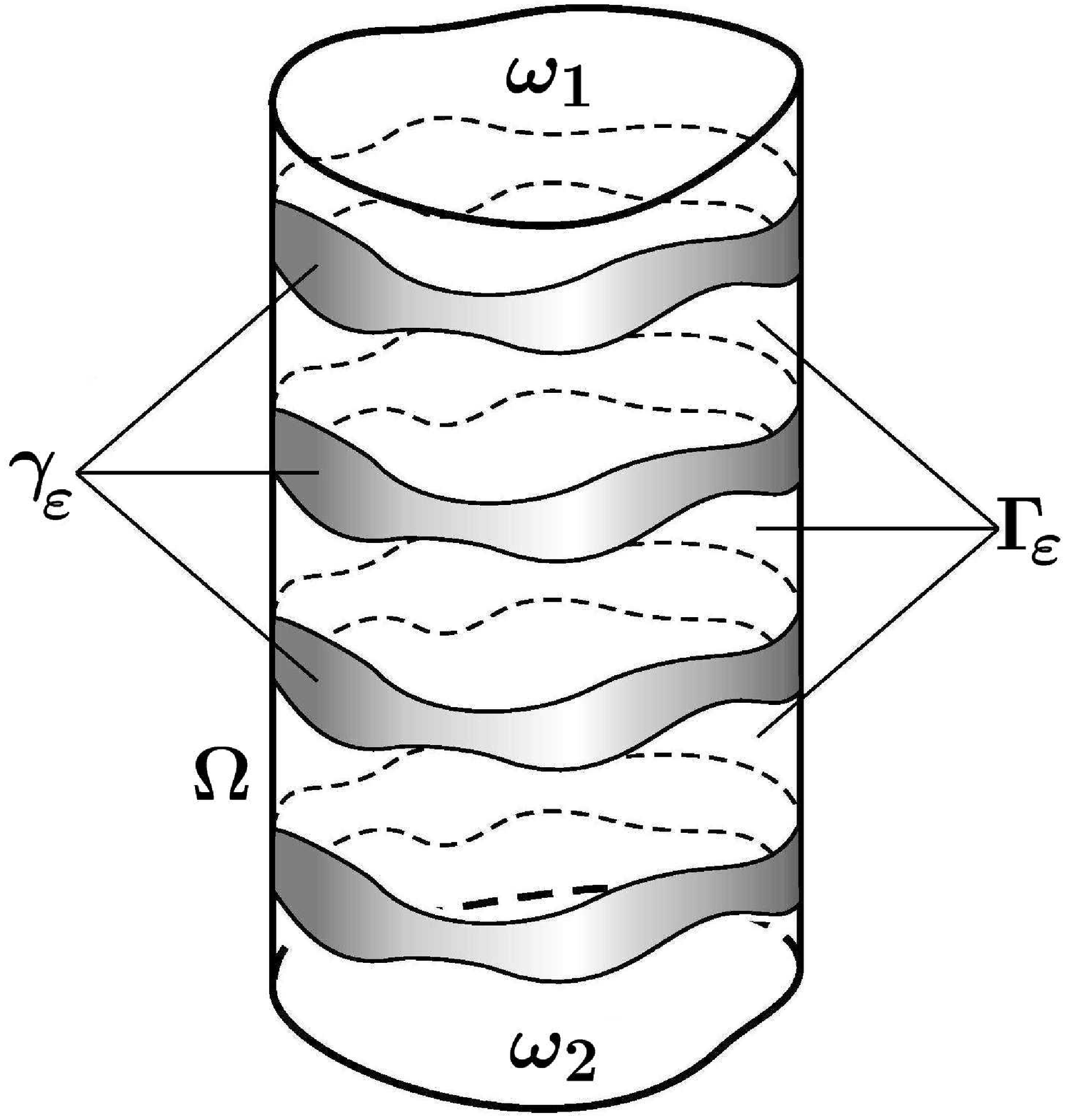}
\end{center}
\end{figure}

In the paper we consider a singular perturbed eigenvalue
problem:
\begin{align}
-&\D\psi_\e=\l_\e\psi_\e,\quad x\in\Om,\label{1.2}
\\
{}&
\begin{aligned}
\psi_\e&=0,\quad x\in\om_1\cup\g_\e,
\\
\frac{\p\psi_\e}{\p\nu}&=0,\quad x\in\om_2\cup\G_\e.
\end{aligned}\label{1.3}
\end{align}
Here $\nu$ is the outward normal for the boundary $\p\Om$, and
the set $\G_\e$ is defined as a complement of $\overline{\g}_\e$
with respect to the lateral surface $\Si$.

Lobo and P\`erez \cite{L-P} studied the homogenization of the
Poisson equation with the boundary condition (\ref{1.3}) for the
case when $\om$ is a unit circle, $\mathsf{g}_\e\equiv 1$. They
established that under the equality
\begin{equation}\label{1.1}
\lim\limits_{\e\to0}\e\ln\eta(\e)=0
\end{equation}
the solution of such problem converges in  $H^1(\Om)$ norm to a
solution of the same Poisson equation with the same boundary
condition on the basis and with the Dirichlet condition on the
lateral surface. For the problem  (\ref{1.2}), (\ref{1.3})
similar statement will be proved in this paper.

\begin{theorem}\label{th1.1}
Suppose the equality (\ref{1.1}) holds. Then eigenvalues of the
perturbed problem converge to eigenvalues of limiting one:
\begin{equation}\label{1.4}
-\D\psi_0=\l_0\psi_0,\quad x\in\Om,\qquad \psi_0=0,\quad
x\in\om_1\cup\Si,\qquad \frac{\p\psi_0}{\p\nu}=0,\quad
x\in\om_2,
\end{equation}
as $\e\to0$. For each eigenfunction $\psi_0$ associated with
eigenvalue $\l_0$ there exists  converging to $\psi_0$  in
$H^1(\Om)$ linear combination of the perturbed eigenfunctions
associated with eigenvalues converging to $\l_0$. Total
multiplicity of the perturbed eigenvalues converging to a same
limiting eigenvalue coincides with the multiplicity of this
limiting eigenvalue.
\end{theorem}

The problem (\ref{1.4}) is easily solved by separation of
variables: $\l_0=M^2+\k$, $\psi_0(x)=\phi_0(x')\cos Mx_3$, where
$M=\pi(m+1/2)H^{-1}$, $m\ge0$ is an integer, $\k$ and $\phi_0$
are eigenelements of two-dimensional problem
\begin{equation}\label{1.7}
-\D_{x'}\phi_0=\k \phi_0,\quad x'\in\om,\qquad \phi_0=0,\quad
x'\in\p\om.
\end{equation}
We arrange the eigenvalues of both perturbed and limiting
problem in ascending order counting multiplicity:
\begin{equation}
\l_0^1\le\l_0^2\le\ldots\le\l_0^k\ldots,\qquad
\l_\e^1\le\l_\e^2\le\ldots\le\l_\e^k\ldots\label{1.10}
\end{equation}
Associated eigenfunctions $\psi_\e^k$ are postulated to be
orthonormalized in $L_2(\Om)$. We denote by $M^k$, $\k^k$ and
$\phi_0^k$ numbers $M$, $\k$ and functions $\phi_0$ associated
with $\l_0^k$. Eigenfunctions of the problem (\ref{1.7}) are
supposed to be orthonormalized in $L_2(\om)$, moreover, the
eigenfunctions associated with multiply eigenvalue are chosen in
such a way their normal derivatives are to be orthogonal in
$L_2(\p\om)$ weighted by $(-\ln\sin\eta \mathsf{g}_\e)$. The
possibility of such orthogonalization follows from well-known
theorem on diagonalization of two quadratic forms in a
finite-dimensional space.

Observe, the problem (\ref{1.4}) can have multiply eigenvalues.
This situation takes place if the problem (\ref{1.7}) has
multiply eigenvalues or for some $i$ and $j$ the equality
$\l_0^i=M_i^2+\k_i^2=M_j^2+\k_j^2=\l_0^j$ holds. Clear, for each
$\k_i$ and $\k_j$ we can always chose the height  $H$ in such
way to achieve the equality $\l_0^i=\l_0^j$.

Let us formulate the main results of the work.
\begin{theorem}\label{th1.2}
Suppose the equality (\ref{1.1}) holds and there exists
$\mathsf{d}>0$ such that a H\"older norm
$\|\mathsf{g}_\e\|_{C^{2+d}(\p\om)}$ is bounded with respect to
$\e$. Then the asymptotics for the eigenvalues of the perturbed
problem have the form:
\begin{align}
\l_\e^k & =\l_0^k+\e\l_1^k(\eta(\e),\e)+
O\left(\e^{3/2}(|\ln\eta|^{3/2}+1)\right),\label{1.5}
\\
\l_1^k(\eta,\e)&=\int\limits_{\p\om}\left(\frac{\p\phi_0^k}{\p
\mathsf{v}}\right)^2\ln\sin\eta
\mathsf{g}_\e\,\mathrm{d}s,\label{1.6}
\end{align}
where $\mathsf{v}$ is outward unit normal for $\p\om$.
\end{theorem}

The statement about the asymptotics of the associated
eigenfunctions under hypothesis of Theorem~\ref{th1.2} will be
formulated in the third section (see Theorem~\ref{th3.1}).

If for some $i\not=j$ the eigenvalues $\l_0^i$ and $\l_0^j$ does
not coincide, then, as it follows from Theorem~\ref{th1.2}, the
eigenvalues $\l_\e^i$ and $\l_\e^j$ does not coincide, too. If
$\l_0^i=\l_0^j$, then for arbitrary domain $\om$ and function
$\mathsf{g}_\e$ the quantities $\l_1^i$ and $\l_1^j$, generally
speaking, are not equal. Thus, in general case the spectrum of
the problem (\ref{1.2}), (\ref{1.3}) consists of simple
eigenvalues only. At the same time, as it was shown in
\cite{BDU}, for a circular cylinder with $\mathsf{g}_\e\equiv 1$
the perturbed problem has also double eigenvalues. It is clear
that even for a circular cylinder with an arbitrary function
$\mathsf{g}_\e$ the perturbed problem, generally speaking, does
not have multiply eigenvalues. In the present paper for the case
of circular cylinder we adduce sufficient conditions for the
function $\mathsf{g}_\e$ under those the perturbed problem has
also multiply eigenvalues; in order to formulate them we
introduce additional notations.

Let $\om$ be a unit circle with center at the origin. Then the
problem  (\ref{1.7}) admits the separation of the variables, its
eigenvalues are roots of equations $\mathcal{J}_n(\sqrt{\k})=0$,
where $\mathcal{J}_n$ are Bessel functions of integer order
$n\ge0$, associated eigenfunctions (not normalized in
$L_2(\Om)$) have the form $\mathcal{J}_0(\sqrt{\k}r)$ ($n=0$),
$\mathcal{J}_n(\sqrt{\k}r)\cos(n\th)$,
$\mathcal{J}_n(\sqrt{\k}r)\sin(n\th)$ ($n>0$), where $(r,\th)$
are polar coordinates, associated with the variables $x'$. All
the roots of the equations $\mathcal{J}_n(\sqrt{\k})=0$ being
distinct \cite{Vat}, the problem (\ref{1.7}) has simple ($n=0$)
and double ($n>0$) eigenvalues only. We continue the function
$\mathsf{g}_\e(\th)$ periodically to all values of  $\th$ by  a
period $2\pi$.

\begin{theorem}\label{th1.3}
Suppose the hypothesis of Theorem~\ref{th1.2} holds, $\om$ is a
unit circle with center at the origin, the function
$\mathsf{g}_\e(\th)$ is periodic on $\th$ over the period
$\pi/(2n)$, $n>0$, $\l_0^k=\k_k^2+M_k^2$ is a double eigenvalue
of the problem (\ref{1.4}), $\k_k$ is a root the equation
$\mathcal{J}_n(\sqrt{\k})=0$. Then the eigenvalue $\l_\e^k$
converging to  $\l_0^k$ is double and has the asymptotics
(\ref{1.5}),
\begin{equation}
\begin{aligned}
\l_1^k(\eta,\e)&=\frac{2\k_k}{\pi}\int\limits_0^{2\pi}
\sin^2(n\th+\a_\e)\ln\sin\eta\mathsf{g}_\e(\th)\,\mathrm{d}\th=
\\
{}&=\frac{2\k_k}{\pi}\int\limits_0^{2\pi}\cos^2(n\th+\a_\e)
\ln\sin\eta\mathsf{g}_\e(\th) \,\mathrm{d}\th,
\end{aligned} \label{1.11}
\end{equation}
where $\a_\e$ is chosen by the constraint
\begin{equation}\label{1.12}
\int\limits_0^{2\pi}\sin(2n\th+2\a_\e)
\ln\sin\eta\mathsf{g}_\e(\th)\,\mathrm{d}\th=0.
\end{equation}
The asymptotics of the associated eigenfunctions are of the form
(\ref{4.3}).
\end{theorem}

The condition  imposed in Theorem~\ref{th1.2} to the function
$\mathsf{g}_\e$, are called to exclude bounded functions
$\mathsf{g}_\e$ having derivatives unbounded on $\e$. By this we
don't deal with rapidly oscillating functions $\mathsf{g}_\e$,
those geometrically corresponds to the strips on the lateral
surface of rapidly varying width. For these cases on the basis
of Theorem~\ref{th1.2} in the paper the degree of convergence
for perturbed eigenvalues are estimated, the result is
formulated in the following theorem.
\begin{theorem}\label{th1.4}
Suppose the equality (\ref{1.1}) holds. Then the estimates
\begin{equation*}
-C_k\e(|\ln\eta|+1)\le \l_\e^k-\l_0^k\le 0,
\end{equation*}
are valid with positive constants $C_k$ independent on $\e$ and
$\eta$.
\end{theorem}

\sect{Convergence of the perturbed eigenelements}

In this section we will prove Theorem~\ref{th1.1} and auxiliary
lemma which will be employed in the proof of
Theorem~\ref{th1.2}.

Throughout this section the eigenvalues of perturbed and
limiting problems are assumed to be arranged in accordance with
(\ref{1.10}), and associated eigenfunctions are supposed to be
orthonormalized in  $L_2(\Om)$. The additional orthogonalization
in $L_2(\p\om)$ for limiting eigenfunctions is not assumed to
take place.

To prove Theorem~\ref{th1.1} we will use

\begin{lemma}\label{lm2.1}
Let $Q$ be an arbitrary compact set in a complex plane
containing no limiting eigenvalues. Then for $\l\in Q$ and  $\e$
sufficiently small the boundary value problem
\begin{equation}
-\D u_\e=\l u_\e+f,\quad x\in\Om, u_\e =0,\quad x\in
\om_1\cup\g_\e,\quad \frac{\p u_\e}{\p\nu}=0,\quad x\in
\om_2\cup\G_\e,\label{2.1}
\end{equation}
is uniquely solvable for each function $f\in L_2(\Om)$ and an
uninform on $\e$, $\eta$, $\l$ and $f$ estimate
\begin{equation}
\|u_\e\|_{H^1(\Om)}\le C\|f\|_{L_2(\Om)}.\label{2.2}
\end{equation}
holds true. The function $u_\e$ converges in $H^1(\Om)$ to the
solution of the problem
\begin{equation}
-\D u_0=\l u_0 +f,\quad x\in\Om,\qquad u_0=0,\quad x\in
\p\Om\bs\overline{\om}_2,\qquad \frac{\p u_0}{\p\nu}=0,\quad
x\in \om_2. \label{2.3}
\end{equation}
uniformly on $\l\in Q$ as  $\e\to0$.
\end{lemma}

\PF{Proof.} Clear, the unique solvability of the problem
(\ref{2.1}) is an implication of the estimate (\ref{2.2}). We
will prove the latter by reductio ad absurdum.  Suppose this
estimate is wrong, then there exist sequences
$\e_k\xrightarrow[k\to\infty]{}0$, $\l_k\in Q$, $f_k\in
L_2(\Om)$, such that for $\e=\e_k$, $\l=\l_k$, $f=f_k$ the
solution of the problem (\ref{2.1}) meets an inequality
\begin{equation}
\|u_{\e_k}\|_{H^1(\Om)}\ge k\|f_k\|_{L_2(\Om)}.\label{2.4}
\end{equation}
Without loss of generality we suppose the function $u_{\e_k}$ is
normalized in  $L_2(\Om)$. Then, multiplying the equation in
(\ref{2.1}) by $u_{\e_k}$ and integrating by parts once we get
that
\begin{equation}
\|u_{\e_k}\|_{H^1(\Om)}\le
C\left(\|u_{\e_k}\|_{L_2(\Om)}+\|f_k\|_{L_2(\Om)}\right)
=C\left(\|f_k\|_{L_2(\Om)}+1\right),\label{2.5}
\end{equation}
where constant $C$ is independent on $k$. From  (\ref{2.4}),
(\ref{2.5}) it follows the boundedness of $u_{\e_k}$ in
$H^1(\Om)$ norm:
\begin{equation}
\|u_{\e_k}\|_{H^1(\Om)}\le C,\label{2.7}
\end{equation}
where constant $C$ is independent on $k$. The inequalities
(\ref{2.7}) and (\ref{2.4}) in an obvious way yield the
convergence in $L_2(\Om)$: $f_k\xrightarrow[k\to\infty]{} 0$.
Next, treating (\ref{2.7}) once again, bearing in mind the
compactness of $Q$ and extracting a subsequence from the
sequence of indexes $k$ if needed, we conclude that $\l_k$
converges to $\l_*\in Q$, and $u_{\e_k}$ converges to $u_*$
weakly in $H^1(\Om)$ and strongly in  $L_2(\Om)$, moreover, the
function  $u_*$ is nonzero due to normalization of  $u_{\e_k}$.
Clear, the function $u_{\e_k}$ vanishes on a set $\{x:
x'\in\p\om, |x_3-\e\pi(j+1/2)|<c\e\eta,
j=\overline{0,N-1}\}\cup\om_1$. Relying on this fact, by analogy
with the proof of Theorem II.4 in \cite{L-P} one can easily show
that $u_*$ vanishes on the lateral surface and the upper basis
of the cylinder  $\Om$. On the other hand, for each function
$v\in H^1(\Om)$, vanishing on the lateral surface and on the
upper basis of the cylinder $\Om$, the obvious integral equality
\begin{equation*}
\int\limits_\Om\left(\nabla u_{\e_k},\nabla
v\right)\,\mathrm{d}x=\int\limits_\Om \left(\l_{k}
u_{\e_k}+f_{k}\right)v\,\mathrm{d}x,
\end{equation*}
takes place, passing in which to a limit as $k\to\infty$, we see
that the function $u_*$ is a nontrivial solution to the problem
\begin{equation*}
-\D u_*=\l_* u_*,\quad x\in\Om,\qquad u_*=0,\quad
x\in\p\Om\bs\overline{\om}_2,\qquad\frac{\p u_*}{\p \nu}=0,\quad
x\in\om_2.
\end{equation*}
Thus, $\l_*\in Q$ is an eigenvalue of the limiting problem, what
contradicts to lemma's hypothesis. The proof of the estimate
(\ref{2.2}) is complete.

Employing now the estimate (\ref{2.2}) instead of (\ref{2.7}),
by similar arguments it is easy to prove a strong in $L_2(\Om)$
and weak in $H^1(\Om)$ convergence of the solution of the
problem (\ref{2.1}) to the solution of the problem (\ref{2.3})
for arbitrary converging sequences: $\e_k\to0$, $\l_k\to\l_*$ as
$k\to\infty$. By this convergence and continuity of $u_0$ on
$\l\in Q$ we deduce an uniform on  $\l$ convergence of $u_\e$ to
$u_0$ (strong in  $L_2(\Om)$ and weak in $H^1(\Om)$). Let us
establish the strong convergence in $H^1(\Om)$. Clear, it is
sufficient to prove the convergence of a norm
$\|u_\e\|_{H^1(\Om)}$ to $\|u_0\|_{H^1(\Om)}$. This fact follows
from obvious assertions
\begin{equation*}
\|u_\e\|^2_{H^1(\Om)}=\l\|u_\e\|_{L_2(\Om)}^2+(u_\e,f)_{L_2(\Om)}
\xrightarrow[\e\to0]{}
\l\|u_0\|_{L_2(\Om)}^2+(u_0,f)_{L_2(\Om)}= \|u_0\|^2_{H^1(\Om)}.
\end{equation*}
The proof is complete.

\smallskip

\PF{\indent Proof of Theorem~\ref{th1.1}.} It is known that the
solutions of the problems (\ref{2.1}) and (\ref{2.3}) are
meromorphic on  $\l$ in the sense of  $H^1(\Om)$ norm, their
singularities are simple poles coinciding with eigenvalues of
perturbed and limiting problems respectively, residua at these
poles are corresponding eigenfunctions.

Let $\l_0=\l_0^{q}=\ldots=\l_0^{q+p-1}$ be a $p$-multiply
eigenvalue of the limiting problem, $p\ge1$, and
$\mathcal{B}_\d(\l_0)$ be a closed circle of radius $\d$ with
center at a point $\l_0$ in a complex plane. We take $\d$
sufficiently small such that the  circle $\mathcal{B}_\d(\l_0)$
contains no limiting eigenvalues except $\l_0$. Then by
analyticity of the solutions to the problems (\ref{2.1}),
(\ref{2.3}) on the parameter $\l$ and Lemma~\ref{lm2.1} we
derive the convergence in  $H^1(\Om)$
\begin{equation}
\frac{1}{2\pi \mathrm{i}}\int\limits_{\p\mathcal{B}_\d}
u_\e\,d\l\xrightarrow[\e\to0]{}\frac{1}{2\pi
\mathrm{i}}\int\limits_{\p\mathcal{B}_\d} u_0\,d\l.\label{2.8}
\end{equation}
Since a circle  $\mathcal{B}_\d(\l_0)$ contains (simple) pole of
the function $u_0$, it follows that the right side of
(\ref{2.8}) is nonzero. Therefore, the left side of (\ref{2.8})
is nonzero, too, i.e., the circle  $\mathcal{B}_\d(\l_0)$
contains (simple) pole of the function $u_\e$. This fact and an
arbitrary choice of  $\d$ immediately imply that the eigenvalues
of the perturbed problem converge to the eigenvalues of the
limiting problem.

Let us establish the convergence of the eigenfunctions. By
direct calculations we check that for
$\l\in\mathcal{B}_\d(\l_0)$, $\l\not=\l_0$, $f=\psi_0^{k}$,
$k=q,\ldots,q+p-1$ the solution to the problem (\ref{2.3}) is a
function
\begin{equation*}
u_0=\frac{\psi_0^{k}}{\l_0-\l}.
\end{equation*}
Substituting this equality into (\ref{2.8}) and calculating
right side, we obtain that left side of (\ref{2.8}) where $u_\e$
is a solution to the problem (\ref{2.1}) with $f=\psi_0^{k}$, is
the needed linear combination converging to $\psi_0^{k}$ in
$H^1(\Om)$.

Let us prove that perturbed eigenvalues $\l_\e^k$,
$k=q,\ldots,q+p-1$, converge to $\l_0$. Suppose that eigenvalues
$\l_\e^j$, $j\in I_0$, converge to $\l_0$. We denote by $l$ the
total multiplicity of all perturbed eigenvalues converging to
$\l_0$; $l=|I_0|$. Showing, that $l=p$, we, clear, will prove
the needed convergence. Since the eigenfunctions $\psi_0^{k}$,
$k=q,\ldots,q+p-1$ are linear independent, the corresponding
linear combinations of the functions $\psi_\e^{j}$, $j\in I_0$,
converging to $\psi_0^{k}$, are linear independent, too. The
functions $\psi_\e^{k}$ are linear independent, therefore, by
Steinitz theorem, the number $l$ can not be less than $p$. On
the other hand, assuming, that $l>p$, by analogy with the proof
of Lemma~\ref{lm2.1} one can show the existence of a sequence
$\e_k\to0$, on which each of  (linear independent) functions
$\psi_\e^{j}$, $j\in I_0$, converges to  a linear combinations
of the functions $\psi_0^{k}$, and also, these combinations are
linear independent. Therefore, the number $p$ does not exceed
$l$, i.e., $l=p$. The proof of Theorem~\ref{th1.1} is complete.

In proving Theorem~\ref{th1.2} we will employ the following
auxiliary statement.

\begin{lemma}\label{lm2.2}
For $\l$ close to $p$-multiply eigenvalue
$\l_0=\l_0^{q}=\ldots=\l_0^{q+p-1}$ the solution of the boundary
value problem (\ref{2.1}) satisfies a representation
\begin{equation}\label{2.9}
u_\e=\sum\limits_{k=q}^{q+p-1}\frac{\psi_\e^k}{\l_\e^{k}-\l}
\int\limits_\Om \psi_\e^{k} f\,\mathrm{d}x+ \widetilde{u}_\e,
\end{equation}
where $\widetilde{u}_\e$ is a holomorphic on $\l$ function
orthogonal to all $\psi_\e^{k}$ in $ L_2(\Om)$,
$k=q,\ldots,q+p-1$. For the functions $\widetilde{u}_\e$ an
uniform on $\e$, $\eta$, $\l$ and $f$ estimate
\begin{equation}
\|\widetilde{u}_\e\|_{H^1(\Om)}\le C\|f\|_{L_2(\Om)}\label{2.10}
\end{equation}
holds true.
\end{lemma}

\PF{Proof.} As it was said in the proof of Theorem~\ref{th1.1},
$u_\e$ is a meromorphic on $\l$ function, having simple poles at
the points  $\l_\e^{k}$, residua at these poles are
corresponding eigenfunctions. Therefore, the equality
\begin{equation}
u_\e=\sum\limits_{k=q}^{q+p-1}
\mathsf{b}_k\frac{\psi_\e^{k}}{\l_\e^{k}-\l} +
\widetilde{u}_\e,\label{2.11}
\end{equation}
is correct, where $\widetilde{u}_\e$ is holomorphic on
$\l\in\mathcal{B}_\d(\l_0)$. We multiply the equation in the
problem (\ref{2.1}) by $\psi_\e^{k}$ and integrate by parts. As
a result we have
\begin{equation*}
(\l^{k}_\e-\l)(\psi_\e^{k},u_\e)=(f,\psi_\e^{k}).
\end{equation*}
Substituting the representation (\ref{2.11}) into the equalities
obtained we deduce:
\begin{equation*}
\mathsf{b}_k=\int\limits_\Om\psi_\e^{k}f\,dx,\qquad
(\widetilde{u}_\e,\psi_\e^{k})=0,
\end{equation*}
what proves (\ref{2.9}). It remains to establish the validity of
the inequality (\ref{2.10}). It is easy to see that $\t u_\e$ is
a solution of the problem (\ref{2.1}) with  right side
\begin{equation*}
f-\sum\limits_{k=q}^{q+p-1}\psi_\e^{k}(f,\psi_\e^{k}),
\end{equation*}
and it is holomorphic on $\l\in\p\mathcal{B}_\d(\l_0)$. That's
why for $\l\in\p\mathcal{B}_\d(\l_0)$ a uniform on $\e$, $\eta$,
$\l$ and $f$ estimate
\begin{equation*}
\|\t u_\e\|_{H^1(\Om)}\le
C\|f-\sum\limits_{k=q}^{q+p-1}\psi_\e^{k}(f,\psi_\e^{k})
\|_{L_2(\Om)}\le C\|f\|_{L_2(\Om)},
\end{equation*}
is valid, which by maximum principle for holomorphic functions
takes place  for  $\l\in\mathcal{B}_\d(\l_0)$, too. The proof is
complete.

\sect{Asymptotics of the perturbed eigenelements}

In this section we will prove Theorem~\ref{th1.2} about
asymptotics of the perturbed eigenvalues, and, under its
hypothesis, Theorem~\ref{th3.1} about asymptotics of associated
eigenfunctions.

\PF{\indent Proof of Theorem~\ref{th1.2}.} We will construct the
asymptotics relying on the method of composite
expansions~\cite{CE} and the multiscaled method~\cite{MS}. Our
strategy is, first, to construct these asymptotics expansions
formally, and, second, to prove rigorously that these expansions
formally constructed do provide asymptotics of the perturbed
eigenelements. It is convenient to distinguish two cases in
formal constructing, depending on whether the limiting or
multiply eigenvalue of the problem (\ref{1.7}) is associated
with the limiting eigenvalue. In formal constructing we will
dwell on the case of simple eigenvalue of the problem
(\ref{1.7}); the case of multiply eigenvalue has just small
differences those will be clarified separately.

We start formal constructing. Let $\l_0=M^2+\k^2$, where $\k$ is
a simple eigenvalue of the problem (\ref{1.7}),
$\psi_0(x)=\phi_0(x')\cos Mx_3$ is the associated eigenfunction,
$\|\phi_0\|_{L_2(\om)}=1$, $\l_\e$ is the perturbed eigenvalue
converging to $\l_0$.

We construct the asymptotics for $\l_\e$ as follows:
\begin{equation}\label{3.1}
\l_\e=\l_0+\e\l_1(\eta,\e).
\end{equation}
The asymptotics for associated eigenfunction is constructed as a
sum of two expansions, outer expansion and boundary layer. Outer
expansion looks as follows:
\begin{equation}\label{3.2}
\psi_\e^{ex}(x,\eta)=(\phi_0(x')+\e\phi_1(x',\eta,\e))\cos Mx_3,
\end{equation}
and  boundary layer is of the form
\begin{equation}\label{3.3}
\psi_\e^{bl}(\xi,s,x_3,\eta)=\e v_1^+(\xi,s,\eta,\e)\cos Mx_3,
\end{equation}
where $\xi=(\xi_1,\xi_2)=(\tau\e^{-1},x_3\e^{-1}-\pi/2)$, $\tau$
is a distance from a point to $\p\om$ measured in the direction
of inward normal. We introduce the boundary layer to satisfy
boundary conditions on $\g_\e$ and $\G_\e$. Moreover, in
constructing of boundary layer we also employ the multiscaled
method, the variable $x_3$ plays ''slow time'' role.

Let us proceed to the constructing of the asymptotics, i.e., to
a determining of the  functions $\l_1$, $\phi_1$, $v_1^+$. First
we substitute (\ref{3.1}) and (\ref{3.2}) into the equation
(\ref{1.2}) and gather the coefficients of the first power of
$\e$. This standard procedure implies the equation for the
function $\phi_1$:
\begin{equation}\label{3.4}
(\D_{x'}+\k)\phi_1=-\l_1\phi_0,\quad x'\in\om.
\end{equation}
The boundary condition for the function $\phi_1$ will be
determined in constructing of the boundary layer. Let us derive
the boundary condition for the function $v_1^+$.  In accordance
with the method of composite expansions we require the sum of
the functions $\psi_\e^{ex}$ and $\psi_\e^{bl}$ to satisfy the
boundary conditions (\ref{1.3}) on $\g_\e$ and $\G_\e$
asymptotically on $\e$. This constraint yields the boundary
conditions for $v_1^+$:
\begin{equation}
v_1^+=-\phi_1^D,\quad \xi\in\g(\eta\mathsf{g}_\e),\qquad\frac{\p
v_1^+}{\p\xi_2}=\phi_0^{\nu},\quad
\xi\in\G(\eta\mathsf{g}_\e),\label{3.5}
\end{equation}
where $\g(a)=\{\xi: \xi_2=0, |\xi_1-\pi j|<a, j\in\mathbb{Z}\}$,
$\G(a)=O\xi_1\bs\overline{\g(a)}$,
\begin{equation*}
\phi_1^D=\phi_1^D(s,\eta,\e)=\phi_1(x',\eta,\e), \quad
\phi_0^{\nu}=\phi_0^{\nu}(s)=\frac{\p}{\p
\mathsf{v}}\phi_0(x'),\quad x'\in\p\om.
\end{equation*}
In order to deduce the equation for the function $v_1^+$, we
first rewrite Laplace operator in the variables $(s,\tau,x_3)$:
\begin{equation}\label{3.7}
\Delta_x=\frac{1}{\mathsf{H}}\left(\frac{\partial}{\partial\tau}
\left(\mathsf{H}\frac{\partial}{\partial\tau}
\right)+\frac{\partial}{\partial
s}\left(\frac{1}{\mathsf{H}}\frac{\partial}{\partial
s}\right)\right)+\frac{\partial^2}{\partial x_3^2},\qquad
\mathsf{H}=1+\tau\mathsf{k},
\end{equation}
$\mathsf{k}=\mathsf{k}(s)=\left(\mathsf{r}''(s),
\mathsf{v}(s)\right)_{\mathbb{R}^2}$,
$\mathsf{v}=\mathsf{v}(s)$, $\mathsf{r}(s)$ is a two-dimensional
vector-function prescribing the curve $\partial\om$,
$\mathsf{k}\in C^\infty(\p\om)$. Now we substitute  (\ref{3.1}),
(\ref{3.3}), (\ref{3.7}) into (\ref{1.2}), go over to the
variables $\xi$ and write out the coefficient of smallest power
of  $\e$. As a result we have the equation for the function
$v_1^+$:
\begin{equation}
\D_\xi v_1^+=0,\quad \xi_2>0.\label{3.8}
\end{equation}
In accordance with the method of composite expansions, we should
construct the solution to the problem (\ref{3.5}), (\ref{3.8}),
decaying exponentially as $\xi_2\to+\infty$.

We will employ the symbol $\mathcal{V}(a)$ for the space of
$\pi$-periodic on $\xi_1$ functions belonging to
$C^\infty(\{\xi: \xi_2>0\}\backslash\{\xi: \xi\not=(\pm a+\pi
j), j\in\mathbb{Z}\})$ and decaying exponentially as
$\xi_2\to+\infty$ uniformly on $\xi_1$ together with all their
derivatives. Denote $\Pi=\{\xi: \xi_2>0, |\xi_1|<\pi/2\}$.

We introduce the function
\begin{equation*}
X(\xi,a)=\mathrm{Re}\,\ln\left(\sin z+\sqrt{\sin^2 z-\sin^2 a
}\right)-\xi_2,
\end{equation*}
$z=\xi_1+\mathrm{i}\xi_2$ is a complex variable. By direct
calculations we check that $X(\xi,a)\in\mathcal{V}(a)\cap
H^1(\Pi)$ is a harmonic in a half-plane $\xi_2>0$  function
being even on $\xi_1$ and satisfying boundary conditions
\begin{equation}\label{3.11}
X=\ln\sin a,\quad x\in\g(a),\qquad \frac{\p X}{\p
\xi_2}=-1,\quad x\in\G(a).
\end{equation}
Thus, the solution of the problem (\ref{3.5}), (\ref{3.8}) is
given by the formula:
\begin{equation}\label{3.30}
v_1^+(\xi,s,\eta,\e)=-\phi_0^{\nu}(s)X(\xi,\eta
\mathsf{g}_\e(s)).
\end{equation}
Then, by virtue of the boundary condition (\ref{3.11}),
\begin{equation*}
v_1^+(\xi,s,\eta,\e)=-\phi_0^{\nu}(s)\ln\sin\eta
\mathsf{g}_\e(s)\quad\text{on}\quad \g(\eta \mathsf{g}_\e(s)).
\end{equation*}
In view of (\ref{3.5}), last equality allows to obtain the
boundary condition for $\phi_1$:
\begin{equation}\label{3.12}
\phi_1=\phi_0^{\nu}\ln\sin\eta\mathsf{g}_\e,\quad x\in\p\om.
\end{equation}
The solvability condition of the boundary value problem
(\ref{3.4}), (\ref{3.12}) is obtained in a standard way: we
multiply both sides of the equation (\ref{3.4}) by $\phi_0$ and
integrate by parts. The equality obtained in this way and
normalization condition for  $\phi_0$ lead us to the formula
(\ref{1.6}).

In order to justify the leading terms of the asymptotics
formally constructed we have to construct additional terms in
the asymptotics  for $\psi_\e$. To the boundary layer one should
add two terms; as a result the boundary layer becomes:
\begin{equation}\label{3.13}
\begin{aligned}
\psi_\e^{bl}(\xi,s,x_3,\eta)=&(\e v_1^+(\xi,s,\eta,\e)+\e^2
v_2^+(\xi,s,\eta,\e))\cos Mx_3+ \\ & +\e^2
v_2^-(\xi,s,\eta,\e)\sin(Mx_3).
\end{aligned}
\end{equation}
The equations for the functions $v_2^\pm$ are got by
substituting of (\ref{3.1}), (\ref{3.7}) and (\ref{3.13}) into
(\ref{1.2}) and writing out the coefficients of the same powers
of $\e$ separately for $\cos (Mx_3)$ and $\sin (Mx_3)$:
\begin{equation}
\D_\xi v_2^+=-\mathsf{k}\frac{\p v_1^+}{\p\xi_2},\quad \D_\xi
v_2^-=2M\frac{\p v_1^+}{\p\xi_1},\quad \xi_2>0.\label{3.14}
\end{equation}
We derive the boundary conditions for $v_2^\pm$ as well as
(\ref{3.5}):
\begin{equation}\label{3.15} \frac{\p
v_2^+}{\p\xi_2}=\phi_1^{\nu},\quad\frac{\p
v_2^-}{\p\xi_2}=0,\quad\xi\in\G(\eta\mathsf{g}_\e),
\end{equation}
where $\phi_1^{\nu}$ is a value of normal derivative of the
function $\phi_1$ on $\p\om$,
$\phi_1^{\nu}=\phi_1^{\nu}(s,\eta,\e)$.

We denote:
\begin{equation*}
Y(\xi,a)=\mathrm{Im}\,\ln\left(\sin z+\sqrt{\sin^2 z-\sin^2 a
}\right)-\frac{\pi}{2}+\xi_1.
\end{equation*}
One can check that $Y\in\mathcal{V}(a)\cap H^1(\Pi)$ is odd on
$\xi_1$ harmonic function together with $X$ satisfying
Cauchy-Riemann conditions:
\begin{equation}\label{3.16}
\frac{\p X}{\p\xi_1}=\frac{\p Y}{\p\xi_2},\quad \frac{\p
X}{\p\xi_2}=-\frac{\p Y}{\p\xi_1}.
\end{equation}
The solutions of the problem (\ref{3.14}), (\ref{3.15}) can be
obtained explicitly:
\begin{equation}\label{3.25}
\begin{aligned}
v_2^+(\xi,s,\eta,\e)=&\frac{1}{2}\mathsf{k}(s)\phi_0^{\nu}(s)
\left(\xi_2 X(\xi,\eta
\mathsf{g}_\e(s))+\int\limits_{\xi_2}^{+\infty} X(\xi_1,t,\eta
\mathsf{g}_\e(s))\,\mathrm{d}t\right)-
\\
{}&-\phi_1^\nu(s,\eta,\e)X((\xi,\eta \mathsf{g}_\e(s)),
\\
v_2^-(\xi,s,\eta,\e)=&-\mathsf{k}(s)M\phi_0^{\nu}(s) \left(\xi_2
Y(\xi,\eta \mathsf{g}_\e(s))+\int\limits_{\xi_2}^{+\infty}
Y(\xi_1,t,\eta \mathsf{g}_\e(s))\,\mathrm{d}t\right).
\end{aligned}
\end{equation}
Clear, $v_2^\pm\in H^1(\Pi)\cap \mathcal{V}(\eta
\mathsf{g}_\e)$. Below we will use following auxiliary lemmas.

It arises from the definition of the set $\mathcal{V}(a)$,
belongings $X,Y\in\mathcal{V}(a)$, evenness $X$ and oddness $Y$
on $\xi_1$
\begin{lemma}\label{lm3.3}
The equalities
\begin{equation*}
\frac{\p X}{\p\xi_1}=0,\quad Y=0,\quad \xi_1=\frac{\pi
k}{2},\quad k\in\mathbb{Z}
\end{equation*}
are true.
\end{lemma}

\begin{lemma}\label{lm3.1}
Suppose function $v\in\mathcal{V}(a)\cap L_2(\Pi)$ satisfies an
equality $\int\limits_{-\pi/2}^{\pi/2}
v(\xi)\,\mathrm{d}\xi_1=0$ for each $\xi_2>0$ and
$\frac{\displaystyle\p v}{\displaystyle\p\xi_1}\in L_2(\Pi)$.
Then an estimate
\begin{equation*}
\|v\|_{L_2(\Pi)}\le \pi\Big\|\frac{\p
v}{\p\xi_1}\Big\|_{L_2(\Pi)}
\end{equation*}
is valid.
\end{lemma}

\PF{Proof.} For $\xi_2>0$ by Poincar\'e inequality we have:
\begin{equation*}
\int\limits_{-\pi/2}^{\pi/2} v^2\,d\xi_1\le \frac{\pi^2}{2}
\int\limits_{\pi/2}^{\pi/2}\left(\frac{\p
v}{\p\xi_1}\right)^2\,\mathrm{d}\xi_1\le\pi^2
\int\limits_{\pi/2}^{\pi/2}\left(\frac{\p
v}{\p\xi_1}\right)^2\,\mathrm{d}\xi_1.
\end{equation*}
Integrating now the inequality obtained over
$\xi_2\in(0,+\infty)$, we arrive at the statement of the lemma.
The proof is complete.

Throughout next lemma we denote by $C$ various nonspecific
constants independent on $a$.

\begin{lemma}\label{lm3.2} As $a\in(0,\pi/2]$ the functions
$X$ and $Y$ posses following properties:
\begin{enumerate}
\def\theenumi{(\arabic{enumi})}

\item\label{lm3.2_it3} For each $\xi_2>0$ the equality
\begin{equation*}
\int\limits_{-\pi/2}^{\pi/2} X(\xi,a)\, \mathrm{d}\xi_1=0
\end{equation*}
holds.

\item\label{lm3.2_it1} The assertions
\begin{align*}
{}&\|X\|_{L_2(\Pi)}=\|Y\|_{L_2(\Pi)}\le C,
\quad\phantom{1^{1\,\,2}} \|\xi_2 X\|_{L_2(\Pi)}=\|\xi_2
Y\|_{L_2(\Pi)}\le \pi\|X\|_{L_2(\Pi)},
\\
{}& \|\nabla_\xi X\|_{L_2(\Pi)}=\sqrt{\pi}|\ln\sin a|^{1/2},
\quad \|\xi_2\nabla_\xi X\|_{L_2(\Pi)}=\|\xi_2\nabla_\xi
Y\|_{L_2(\Pi)}=\|X\|_{L_2(\Pi)},
\\
{}& \Big\|\int\limits_{\xi_2}^{+\infty}
X(\xi_1,t,a)\,\mathrm{d}t\Big\|_{L_2(\Pi)}=
\Big\|\int\limits_{\xi_2}^{+\infty}
Y(\xi_1,t,a)\,\mathrm{d}t\Big\|_{L_2(\Pi)}\le\pi
\|X\|_{L_2(\Pi)}
\end{align*}
are true.

\item\label{lm3.2_it4} For functions
$\frac{\displaystyle \p X}{\displaystyle\p
a},\frac{\displaystyle\p Y}{\displaystyle\p a}\in
\mathcal{V}(a)\cap L_2(\Pi)$,
\begin{equation*}
\frac{\p} {\p a}\int\limits_{\xi_2}^{+\infty}
X(\xi_1,t,a)\,\mathrm{d}t,\; \frac{\p} {\p
a}\int\limits_{\xi_2}^{+\infty}
X(\xi_1,t,a)\,\mathrm{d}t\in\mathcal{V}(a)\cap H^1(\Pi)
\end{equation*}
the assertions
\begin{align*}
{}&\Big\|\frac{\p X}{\p a}\Big\|_{L_2(\Pi)}= \Big\|\frac{\p
Y}{\p a}\Big\|_{L_2(\Pi)}=\frac{\sqrt{\pi}\cot a|\ln\cos
a|^{1/2}}{\sqrt{2}},
\\
{}& \Big\|\xi_2\frac{\p X}{\p a}\Big\|_{L_2(\Pi)}=
\Big\|\xi_2\frac{\p Y}{\p a}\Big\|_{L_2(\Pi)} \le\pi
\Big\|\frac{\p X}{\p a}\Big\|_{L_2(\Pi)},
\\
{}&\Big\|\frac{\p}{\p a}\int\limits_{\xi_2}^{+\infty}
X(\xi_1,t,a)\,\mathrm{d}t\Big\|_{L_2(\Pi)}= \Big\|\frac{\p}{\p
a}\int\limits_{\xi_2}^{+\infty}
Y(\xi_1,t,a)\,\mathrm{d}t\Big\|_{L_2(\Pi)} \le\pi \Big\|\frac{\p
X}{\p a}\Big\|_{L_2(\Pi)}
\end{align*}
hold.
\end{enumerate}
\end{lemma}

\PF{Proof.} Throughout the proof, not saying it specially, in
various integrating by parts we will employ the boundary
conditions for  $X$ and $Y$ from Lemma~\ref{lm3.3}.

The statement of the item \ref{lm3.2_it3} can be easily obtained
by integrating by parts in equalities ($t>0$)
\begin{equation*}
\int\limits_{\Pi\cap\{\xi: \xi_2>t\}}\D_\xi X\,\mathrm{d}\xi=0,
\quad \int\limits_{\Pi\cap\{\xi: \xi_2>t\}}\xi_2\D_\xi
X\,\mathrm{d}\xi=0.
\end{equation*}
We proceed to the proof the items \ref{lm3.2_it1},
\ref{lm3.2_it4}. The belongings
\begin{equation*}
\frac{\p X}{\p a},\; \frac{\p Y}{\p a},\; \frac{\p}{\p a}
\int\limits_{\xi_2}^{+\infty} X(\xi_1,t,a)\,\mathrm{d}t,\;
\frac{\p} {\p a}\int\limits_{\xi_2}^{+\infty}
X(\xi_1,t,a)\,\mathrm{d}t\in \mathcal{V}(a)\cap L_2(\Pi)
\end{equation*}
are established relying on the explicit form of $X$ and $Y$. The
derivatives of these functions on $\xi_1$, $\xi_2$ equal  to the
functions $\frac{\displaystyle\p X}{\displaystyle\p a}$,
$\frac{\displaystyle\p Y}{\displaystyle\p a}$ due to
(\ref{3.16}), what proves the belongings to a space
$\mathcal{V}(a)\cap H^1(\Pi)$ for the functions
$\frac{\displaystyle\p}{\displaystyle\p a}
\int\limits_{\xi_2}^{+\infty} X(\xi_1,t,a)\,\mathrm{d}t$,
$\frac{\displaystyle\p}{\displaystyle\p
a}\int\limits_{\xi_2}^{+\infty} X(\xi_1,t,a)\,\mathrm{d}t$. The
existence of other norms from the items \ref{lm3.2_it1},
\ref{lm3.2_it4} follows from the explicit forms of the functions
$X$ and $Y$. Let us prove the coincidence of the corresponding
norms of  $X$ and $Y$ from \ref{lm3.2_it1}, \ref{lm3.2_it4}.
Using Cauchy-Riemann conditions (\ref{3.16}) and integrating by
parts, for $\xi_2>0$ we get:
\begin{align*}
\frac{\p}{\p\xi_2}\int\limits_{\pi/2}^{\pi/2}
Y^2\,\mathrm{d}\xi_1=&2\int\limits_{\pi/2}^{\pi/2} Y\frac{\p Y
}{\p\xi_2}\,\mathrm{d}\xi_1=2\int\limits_{\pi/2}^{\pi/2}
Y\frac{\p X}{\p\xi_1}\,\mathrm{d}\xi_1=
\\
=&-2\int\limits_{\pi/2}^{\pi/2} \frac{\p
Y}{\p\xi_1}X\,\mathrm{d}\xi_1=2\int\limits_{\pi/2}^{\pi/2}
\frac{\p X}{\p\xi_2}X\,\mathrm{d}\xi_1
=\frac{\p}{\p\xi_2}\int\limits_{\pi/2}^{\pi/2}
X^2\,\mathrm{d}\xi_1,
\end{align*}
from what it follows that
\begin{equation*}
\int\limits_{\pi/2}^{\pi/2} Y^2\,\mathrm{d}\xi_1=
\int\limits_{\pi/2}^{\pi/2} X^2\,\mathrm{d}\xi_1,\quad \xi_2>0.
\end{equation*}
The equality obtained proves that
$\|X\|_{L_2(\Pi)}=\|Y\|_{L_2(\Pi)}$, $\|\xi_2
X\|_{L_2(\Pi)}=\|\xi_2 Y\|_{L_2(\Pi)}$. The coincidence of other
norms for $X$ and $Y$ is established by analogy.

We proceed to the proof of the estimates and other equalities
from the items \ref{lm3.2_it1}, \ref{lm3.2_it4}. In \cite[\S
3]{MZ} it was shown that $\|X\|_{L_2(\Pi)}$ is continuous on
$a\in[0,\pi/2]$ function, from what it follows the needed
estimate for this function. In \cite{AA} it was proved that
\begin{equation}\label{3.20}
\int\limits_{\gamma(a)\cap\overline{\Pi}}\frac{\partial
X}{\partial\xi_2}\,d\xi_1=\pi-2a,\quad
\int\limits_{\Gamma(a)\cap\overline{\Pi}} X\,d\xi_1=-2a\ln\sin
a.
\end{equation}
Integrating by parts in an equality
$\int\limits_{\Pi}X\Delta_\xi X\,d\xi=0$, we obtain:
\begin{align*}
\int\limits_\Pi\left|\nabla_\xi X\right|^2\,d\xi=-\ln\sin a
\int\limits_{\gamma(a)\cap\overline{\Pi}}\frac{\partial
X}{\partial\xi_2}\,d\xi_1+
\int\limits_{\Gamma(a)\cap\overline{\Pi}}X\,d\xi_1,
\end{align*}
from what and (\ref{3.20}) it follows the maintained formula for
$\|\nabla_\xi X\|_{L_2(\Pi)}$. Equalities
\begin{equation}\label{3.26}
\begin{aligned}
0&=\int\limits_{\Pi}\xi_2^2 X\Delta_\xi
X\,d\xi=-\int\limits_{\Pi} \xi_2^{2}\Big|\nabla_\xi
X\Big|^2\,d\xi-2\int\limits_{\Pi}\xi_2 X \frac{\partial
X}{\partial\xi_2}\,d\xi=
\\
{}&=-\| \xi_2\nabla_\xi X\|_{L_2(\Pi)}^2+\| X\|_{L_2(\Pi)}^2
\end{aligned}
\end{equation}
imply needed expression for $\|\xi_2\nabla_\xi X\|_{L_2(\Pi)}$.
By Lemma~\ref{lm3.1} and the item \ref{lm3.2_it3} we deduce:
\begin{equation*}
\|\xi_2 X\|_{L_2(\Pi)}\le \pi\Big\|\xi_2\frac{\p
X}{\p\xi_1}\Big\|_{L_2(\Pi)}\le \pi\|\xi_2\nabla_\xi
X\|_{L_2(\Pi)}=\pi \| X\|_{L_2(\Pi)}.
\end{equation*}
Basing on the item \ref{lm3.2_it3}, Lemma~\ref{lm3.1},
(\ref{3.16}) and the proven equalities and estimates from the
item \ref{lm3.2_it1},  we establish that
\begin{align*}
\Big\|\int\limits_{\xi_2}^{+\infty}
X(\xi_1,t,a)\,\mathrm{d}t\Big\|_{L_2(\Pi)}\le\pi
\Big\|\int\limits_{\xi_2}^{+\infty} \frac{\p
X}{\p\xi_1}(\xi_1,t,a)\,\mathrm{d}t\Big\|_{L_2(\Pi)}=\pi
\|Y\|_{L_2(\Pi)}=\pi \|X\|_{L_2(\Pi)}.
\end{align*}
The proof of the item \ref{lm3.2_it1} is complete. By direct
calculations one can  easily check that
\begin{equation*}
X_1(\xi,a)=-\frac{1}{2}\xi_2 \int\limits_{\xi_2}^{+\infty}
\frac{\p X}{\p a}(\xi_1,t,a)\,\mathrm{d}t\in
H^1(\Pi)\cap\mathcal{V}(a)
\end{equation*}
is an even on $\xi_1$ solution to a problem
\begin{equation}\label{3.29}
\begin{aligned}
{}&\D_\xi X_1(\xi)=\frac{\p X}{\p a},\quad \xi_2>0,
\\
X_1=0,\quad &\frac{\p
X_1}{\p\xi_2}=-\frac{1}{2}\int\limits_0^{+\infty} \frac{\p X}{\p
a} (\xi_1,t,a)\,\mathrm{d}t,\quad \xi_2=0.
\end{aligned}
\end{equation}
Since for $\xi_1\in(a,\pi/2]$
\begin{equation*}
\frac{\p^2}{\p\xi_1^2}\int\limits_{0}^{+\infty} \frac{\p X}{\p
a}(\xi_1,t,a)\,\mathrm{d}t=-
\int\limits_{0}^{+\infty}\frac{\p^3}{\p t^2\p a
}X(\xi_1,t,a)\,\mathrm{d}t=0,
\end{equation*}
in view of evenness and $\pi$-periodicity on $\xi_1$ of the
function $X$ we derive ($\xi_1\in(a,\pi/2]$):
\begin{equation*}
\frac{\p}{\p a}\int\limits_{0}^{+\infty}
X(\xi_1,t)\,\mathrm{d}t=\int\limits_{0}^{+\infty} \frac{\p X
}{\p a}(\pi/2,t,a)\,\mathrm{d}t=\cot a\ln\cos a.
\end{equation*}
Relying on the statement of the item \ref{lm3.2_it3},
integrating by parts and bearing in mind last equality and
(\ref{3.29}) we get:
\begin{align*}
{}&\int\limits_{\Pi}\left(\frac{\p X}{\p
a}\right)^2\,\mathrm{d}\xi =\int\limits_{\Pi}\left(\frac{\p
X}{\p a}-\cot a\right)\frac{\p X}{\p a}\,\mathrm{d}\xi=
\int\limits_{\Pi}\left(\frac{\p X}{\p a}-\cot
a\right)\D_\xi\frac{\p X_1}{\p a}\,\mathrm{d}\xi=
\\
{}&=\int\limits_{a}^{\pi/2} \left(\frac{\p}{\p
a}X(\xi_1,0,a)-\cot a\right)\frac{\p}{\p a}
\int\limits_0^{+\infty}X(\xi_1,\xi_2,a)\,
\mathrm{d}\xi_2\mathrm{d}\xi_1=
\\
{}&=\cot\,a\ln\cos a\int\limits_{a}^{\pi/2} \left(\frac{\p}{\p
a}X(\xi_1,0,a)-\cot a\right)\,
\mathrm{d}\xi_1=-\frac{\pi}{2}\cot^2\,a\ln\cos a.
\end{align*}
By analogy with (\ref{3.26}) we prove the equality
\begin{equation*}
\Big\|\xi_2\nabla_\xi\frac{\p X}{\p a}\Big\|_{L_2(\Pi)}
=\Big\|\frac{\p X}{\p a}\Big\|_{L_2(\Pi)},
\end{equation*}
what together with the estimate
\begin{equation*}
\Big\|\frac{\p X}{\p a}\Big\|_{L_2(\Pi)}\le\pi \Big\|\frac{\p^2
X}{\p\xi_1\p a}\Big\|_{L_2(\Pi)}\le\pi
\Big\|\xi_2\nabla_\xi\frac{\p X}{\p a}\Big\|_{L_2(\Pi)}
\end{equation*}
implied by  \ref{lm3.2_it3} and Lemma~\ref{lm3.1} lead to the
second estimate of the item \ref{lm3.2_it4}. Third estimate is
established on the base of  \ref{lm3.2_it3}, Lemma~\ref{lm3.1}
and Cauchy-Riemann conditions (\ref{3.16}):
\begin{equation*}
\Big\|\frac{\p}{\p a}\int\limits_{\xi_2}^{+\infty}
X(\xi_1,t,a)\,\mathrm{d}t\Big\|_{L_2(\Pi)}\le\pi\Big\|\frac{\p
Y}{\p a}\Big\|_{L_2(\Pi)}=\pi\Big\|\frac{\p X}{\p
a}\Big\|_{L_2(\Pi)}.
\end{equation*}
The proof is complete.

\begin{lemma}\label{lm3.4}
Suppose that there exists $\mathsf{d}>0$, such that a H\"older
norm $\|\mathsf{g}_\e\|_{C^{2+d}(\p\om)}$ is bounded on $\e$.
Then an uniform on $\e$ and $\eta$ estimate
\begin{equation*}
\|\phi_1\|_{C^2(\overline{\om})}\le C(|\ln\eta|+1)
\end{equation*}
is correct.
\end{lemma}

\PF{Proof.} From (\ref{1.6}) and normalization condition for
$\phi_0$ it obviously arises an uniform on $\e$ and $\eta$
estimate:
\begin{equation}\label{3.27}
|\l_1|\le C(|\ln\eta|+1).
\end{equation}
Therefore, in accordance with general theory of elliptic
boundary value problems and theorem on embedding
$H^2(\om)\subset C(\overline{\om})$ and due to orthogonality
$\phi_1$ and $\phi_0$ an inequality
\begin{equation*}
\|\phi_1\|_{C(\overline{\om})}\le C\|\phi_0\|_{H^2(\om)}\le
C\left(|\l_1|\|\phi_1\|_{L_2(\om)}+
\|\phi_0^\nu\ln\sin\eta\mathsf{g}_\e\|_{C^2(\p\om)}\right)
\le
C(|\ln\eta|+1)
\end{equation*}
takes place. Employing this estimate for
$C(\overline{\om})$-norm of  $\phi_1$ and Schauder inequalities
\cite[Chapter III, \S 1, formula (1.11)]{Ld}, we deduce that
\begin{equation*}
\|\phi_1\|_{C^{2+d}(\overline{\om})}\le C\left(|\l_1|
\|\phi_0\|_{C^{d}(\overline{\om})}+
\|\phi_1\|_{C(\overline{\om})}+\|\phi_0^\nu\ln\sin\eta
\mathsf{g}_\e\|_{C^{2+d}(\overline{\om})}\right),
\end{equation*}
from what and  (\ref{3.27}) the statement of the lemma follows.
The proof is complete.

Let $\chi(t)$ be an infinitely differentiable cut-off function,
equalling to one as $t<1/4$ and vanishing as $t>3/4$, $c_0$ is a
sufficiently small fixed positive number such that in a domain
$\{x': |\tau|<c_0\}$ the variables $(s,\tau)$ are defined
correctly. We denote:
\begin{equation*}
\widetilde{\psi}_\e^{bl}(x,\eta)=\e^2(v_2^+(\xi,s,\eta,\e)\cos
Mx_3 +v_2^-(\xi,s,\eta,\e)\sin(Mx_3))\chi(\tau/c_0).
\end{equation*}
From the definition of the functions $v_2^\pm$ and
Lemmas~\ref{lm3.3},~\ref{lm3.2},~\ref{lm3.4} it follows

\begin{lemma}\label{lm3.5}
The function $\widetilde{\psi}_\e^{bl}\in H^1(\Om)\cap C^\infty(
\overline{\Om}\backslash(\overline{\g}_\e\cap
\overline{\G}_\e))$ satisfies boundary conditions
\begin{equation*}
\widetilde{\psi}_\e^{bl}=0,\quad x\in\om_1,\qquad
\frac{\p}{\p\nu}\widetilde{\psi}_\e^{bl}=0,\quad
x\in\om_2,\qquad
\frac{\p}{\p\nu}\widetilde{\psi}_\e^{bl}=0,\quad x\in\g_\e.
\end{equation*}
Under hypothesis of Lemma~\ref{lm3.4} uniform on $\e$ and $\eta$
estimates
\begin{align*}
{}&\|\widetilde{\psi}_\e^{bl}\|_{H^1(\Om)}\le
C\e^{3/2}\left(|\ln\eta|^{1/2}+1\right),
\\
{}&\|\psi_\e^{bl}\|_{H^1(\Om)}\le C\e^{1/2}\left(
|\ln\eta|^{1/2}+1\right),
\\
{}&\Big\|\chi(\tau/c_0)\frac{\p\psi_\e^{bl}}{\p
s}\Big\|_{L_2(\Om)}\le C\e^{3/2}
\end{align*}
holds true.
\end{lemma}

\begin{lemma}\label{lm3.6}
Suppose the hypothesis of Lemma~\ref{lm3.4} holds. Then there
exists a solution $\psi_2\in H^1(\Om)\cap C^\infty(
\overline{\Om}\backslash(\overline{\g}_\e\cup
\overline{\G}_\e))$ to the boundary value problem
\begin{equation}\label{3.28}
\begin{aligned}
{}&(\D-1)\psi_2=-\e^2\l_1\psi_1-\frac{\chi(\tau/c_0)}{\mathsf{H}}\frac{\p}{\p
s} \left(\frac{1}{\mathsf{H}}\frac{\p\psi_\e^{bl}}{\p s}\right),
\quad x\in\Om,
\\
{}&\psi_2=0,\quad x\in\om_1,\qquad
\psi_2=-\widetilde{\psi}_\e^{bl},\quad x\in\g_\e,\qquad
\frac{\p\psi_2}{\p\nu}=0,\quad x\in\om_2\cup\G_\e.
\end{aligned}
\end{equation}
This solution meets an uniform on $\e$ and $\eta$ estimate
\begin{equation*}
\|\psi_2\|_{H^1(\Om)}\le
C\e^{3/2}\left(\e^{1/2}|\ln\eta|^2+|\ln\eta|^{1/2}+1\right).
\end{equation*}
\end{lemma}

\PF{Proof.} Following \cite{Ld}, by a solution of the problem
(\ref{3.28}) we mean a solution of an integral equation
\begin{equation*}
-(\psi_2,v)_{H^1(\Om)}=-\e^2\l_1(\psi_1,v)_{L_2(\Om)}+
\left(\frac{\chi(\tau/c_0)}{\mathsf{H}}\frac{\p\psi_\e^{bl}}{\p
s}, \frac{1}{\mathsf{H}}\frac{\p v}{\p s}\right)_{L_2(\Om)},
\end{equation*}
whose trace on $\om_1$ is zero and  trace on $\g_\e$ equals to
$\widetilde{\psi}_\e^{bl}$, where $v\in
H^1(\Om;\g_\e\cup\om_1)\equiv\{v: v\in H^1(\Om),
v=0\;\;\text{on}\;\; \g_\e\cup\om_1\}$. The right side of this
integral equality is estimated above by a quantity
\begin{equation*}
C\left(\e^2|\l_1|\|\phi_1\|_{L_2(\om)}+ \Big\|\chi(\tau/c_0)
\frac{\p\psi_\e^{bl}}{\p
s}\Big\|_{L_2(\Om)}\right)\|v\|_{H^1(\Om)},
\end{equation*}
where $C$ is independent on $\e$, $\eta$, $\l_1$, $\phi_1$, and
$v$. By virtue of this estimate,  following the ideas of
\cite{Ld}, one can easily prove the existence of the solution to
(\ref{3.28}) in $H^1(\Om)$ and an inequality
\begin{equation*}
\|\psi_2\|_{H^1(\Om)}\le C\left(\e^2|\l_1|\|\phi_1\|_{L_2(\om)}+
\Big\|\chi(\tau/c_0) \frac{\p\psi_\e^{bl}}{\p
s}\Big\|_{L_2(\Om)}\right).
\end{equation*}
The inequality obtained due to (\ref{3.27}) and
Lemmas~\ref{lm3.4},~\ref{lm3.5} yields the maintained estimate
for $\|\psi_2\|_{H^1(\Om)}$. The belonging
$\|\psi_2\|_{H^1(\Om)} \in C^\infty(
\overline{\Om}\backslash(\overline{\g}_\e\cap
\overline{\G}_\e))$ is established by the theorems on the
smoothness of solutions to elliptic boundary value problems. The
proof is complete.

We set:
\begin{equation*}
\widehat{\l}_\e=\l_0+\e\l_1(\e,\eta), \quad
\widehat{\psi}_\e(x)=\psi_\e^{ex}(x,\eta)+\chi(\tau/c_0)
\psi_\e^{bl}(\xi,s,x_3,\eta)+\psi_2(x,\eta,\e).
\end{equation*}
Next lemma maintains that formally constructed asymptotics for
the eigenelements are formal asymptotics solution of the
perturbed problem.

\begin{lemma}\label{lm3.7}
Suppose the hypothesis of Theorem~\ref{th1.2} holds. Then
functions $\widehat{\l}_\e$ and $\widehat{\psi}_\e\in
H^1(\Om)\cap C^\infty(
\overline{\Om}\backslash(\overline{\g}_\e\cap
\overline{\G}_\e))$ satisfy the boundary value problem
(\ref{2.1}) with $u_\e=\widehat{\psi}_\e$, $\l=\widehat{\l}_\e$,
$f=f_\e$, where for $f_\e$ an uniform on $\e$ and $\eta$
estimate
\begin{equation*}
\|f_\e\|_{L_2(\Om)}\le C\e^{3/2}(|\ln\eta|^{3/2}+1).
\end{equation*}
holds. The equalities $\widehat{\l}_\e=\l_0+o(1)$,
$\|\widehat{\psi}_\e-\psi_0\|_{H^1(\Om)}=o(1)$ are correct  as
$\e\to0$.
\end{lemma}

\PF{Proof.} The convergence of $\widehat{\l}_\e$ to $\l_0$
follows from the estimate (\ref{3.27}), and the equality
$\|\widehat{\psi}_\e-\psi_0\|_{H^1(\Om)}=o(1)$  does from
Lemmas~\ref{lm3.4}-\ref{lm3.6}. Boundary conditions for
$\widehat{\psi}_\e$ follows from (\ref{3.5}), (\ref{3.11}),
(\ref{3.12}), (\ref{3.13}), (\ref{3.15}) and
Lemmas~\ref{lm3.3},~\ref{lm3.6}. Due to  (\ref{1.7}),
(\ref{3.4}), (\ref{3.7}) and (\ref{3.28}) the function
$f_\e=-(\D+\widehat{\l}_\e)\widehat{\psi}_\e$ meets a
representation:
\begin{align*}
{}&f_\e=-\sum\limits_{i=1}^3f_\e^{(i)},\quad
f_\e^{(1)}=(\widehat{\l}_\e+1)\widehat{\psi}_\e,
\\
{}&f_\e^{(2)}=\chi(\tau/c_0)\left(\frac{1}{\mathsf{H}}
\frac{\p}{\p\tau}\left(\mathsf{H}\frac{\p}{\p\tau}\right)+
\frac{\p^2}{\p x_3^2}+\widehat{\l}_\e \right)\psi_\e^{bl},
\\
{}&f_\e^{(3)}=2(\nabla\chi(\tau/c_0),
\nabla\psi_\e^{bl})_{\mathbb{R}^3}+
\psi_\e^{bl}\D\chi(\tau/c_0).
\end{align*}
We obtain from Lemma~\ref{lm3.6} and (\ref{1.1}), (\ref{3.27})
that
\begin{equation*}
\|f_\e^{(1)}\|_{L_2(\Om)}\le C\e^{3/2}(|\ln\eta|^{3/2}+1),
\end{equation*}
where $C$ is independent on $\e$ and $\eta$. Employing
Lemmas~\ref{lm3.2},~\ref{lm3.4}, the equality (\ref{1.1}),
equations (\ref{3.8}), (\ref{3.14}), formulae (\ref{3.30}),
(\ref{3.25}), Cauchy-Riemann conditions (\ref{3.16}) and an
estimate $\e\xi_2<c_0$ that is valid in a domain $\{x':
\tau<c_0\}$, we get
\begin{equation*}
\|f_\e^{(2)}\|_{L_2(\Om)}\le C\e^{3/2}(|\ln\eta|^{1/2}+1),
\end{equation*}
where $C$ is independent on $\e$ and $\eta$. Belongings $X,Y\in
\mathcal{V}(a)$, the definitions of  $\chi$ and explicit
definition of $X$ and $Y$ imply
\begin{equation*}
\|f_\e^{(3)}\|_{L_2(\Om)}\le C\mathrm{e}^{-1/\e^\mathsf{b}},
\end{equation*}
where $\mathsf{b}>0$ is a some fixed number, $C$ is independent
on $\e$ and $\eta$. Gathering together the inequalities for
$f_\e^{(i)}$ obtained, we arrive at the maintained estimate for
$f_\e$. The proof is complete.

Formal constructing in the case of multiply eigenvalue $\k$
actually does not differ from one given above almost in all
details. Here we simultaneously asymptotics of several
eigenelements. The condition of additional orthogonalization in
$L_2(\p\om)$ for the eigenfunctions of the problem (\ref{1.7})
associated with multiply eigenvalue described in the first
section is a solvability condition of the problem (\ref{3.4}),
(\ref{3.12}). All other arguments are not needed to be changed
and are independent   on the multiplicity of  $\k$. Thus, in the
case of $p$-multiply eigenvalue $\k=\k_q=\ldots=\k_{q+p-1}$ as a
result of constructing we have $2p$ functions $\h\l_\e^k$ and
$\h\psi_\e^k$, corresponding to $\k_k$, $\phi_0^k$ and defined
as well as  $\widehat{\l}_\e$ and $\widehat{\psi}_\e$. The
functions $\h\l_\e^k$ and $\h\psi_\e^k$ obey Lemma~\ref{lm3.7},
we indicate by $f_\e^k$ the function $f_\e$ from this lemma
associated with  $\h\l_\e^k$ and $\h\psi_\e^k$.

We start the justification of the asymptotics. Suppose
$\l_0=\l_0^q=\ldots=\l_0^{q+p-1}$ is $p$-multiply limiting
eigenvalue, $p\ge1$. Due to lemmas~\ref{lm2.2}~and~\ref{lm3.7}
the functions $\h\psi_\e^k$ meet the representations
($k=q,\ldots,q+p-1$):
\begin{align}
\h\psi_\e^k&=\sum\limits_{i=q}^{q+p-1}\mathsf{b}^\e_{ki}
\psi_\e^i+\t u_\e^k,\label{3.31}
\\
\mathsf{b}^\e_{ki}&=\frac{1}{\l_\e^i-\h\l_\e^k}\int\limits_{\Om}
\psi_\e^i f_\e^k\,\mathrm{d}x,\label{3.32}
\end{align}
where a function $\t u_\e^k$ is orthogonal to the eigenfunctions
$\psi_\e^i$, $i=q,\ldots,q+p-1$, in  $L_2(\Om)$ and satisfies an
uniform on $\e$ and $\eta$ estimate:
\begin{equation}\label{3.33}
\|\t u_\e^k\|_{H^1(\Om)}\le C\e^{3/2}(|\ln\eta|^{3/2}+1).
\end{equation}
Now we multiply the representation (\ref{3.31}) by $\psi_\e^i$
in $L_2(\Om)$ and bear in mind the orthonormalization condition
for $\psi_\e^k$ and orthogonality of $\psi_\e^i$ and $\t
u_\e^k$, then we have:
\begin{equation}\label{3.34}
\mathsf{b}^\e_{ki}=(\h\psi_\e^k,\psi_\e^i)_{L_2(\Om)}.
\end{equation}
Let us prove the correctness of the asymptotics (\ref{1.5}) by
reductio ad absurdum. Suppose that at some sequence $\e_j\to0$
some of eigenvalues $\l_\e^k$, $k=q,\ldots,q+p-1$, does not
satisfy the asymptotics (\ref{1.5}), i.e., the inequalities
\begin{equation*}
|\l_\e^i-\h\l_\e^k|\ge
j(\e_j^{3/2}(|\ln\eta(\e_j)|^{3/2}+1)),\qquad i\in I_0,\quad
k=q,\ldots,q+p-1,
\end{equation*}
hold, where $I_0\subseteq\{q,\ldots,q+p-1\}$ is a subset of
indexes of the perturbed eigenvalues not satisfying the
maintained asymptotics. From these estimates, (\ref{3.32}) and
the estimates for  $f^k_\e$ from Lemma~\ref{lm3.7} we get:
\begin{equation*}
|\mathsf{b}^{\e_j}_{ki}|\le
C/j\xrightarrow[j\to\infty]{}0,\qquad i\in I_0,\quad
k=q,\ldots,q+p-1.
\end{equation*}
From (\ref{3.34}) and Lemma~\ref{lm3.7} the boundedness of
$\mathsf{b}^\e_{ki}$ follows, that's why, extracting a
subsequence  from $\e_j$ if needed, we assume that
$\mathsf{b}^{\e_j}_{ki}\to \mathsf{b}^0_{ki}$ as $j\to\infty$,
$k,i=q,\ldots,q+p-1$, moreover, as it has been established,
$\mathsf{b}^0_{ki}=0$, $k=q,\ldots,q+p-1$, $i\in I_0$. By the
numbers $\mathsf{b}^{\e}_{ki}$ we compose  $p$ vectors
$\mathsf{b}^{\e}_{k}$ following a rule: a vector
$\mathsf{b}^{\e}_{k}$ consists of numbers
$\mathsf{b}^{\e}_{ki}$, we the index  $i$ consequently takes the
values from the set $\{q,\ldots,q+p-1\}\backslash I_0$. By
analogy we compose vectors $\mathsf{b}^0_k$. Clear, the
dimensions of these vectors are $(p-|I_0|)<p$ and the
convergences $\mathsf{b}^{\e_j}_k\to\mathsf{b}^{0}_k$  hold.
Taking into account this convergence, we multiply the
representations (\ref{3.31})  each by other in $L_2(\Om)$ and
employ the orthonormalization condition for  $\psi_\e^k$ and
$\psi_0^k$, Lemma~\ref{lm3.7}, and the estimate (\ref{3.33});
then we have:
\begin{equation*}
(\mathsf{b}^0_k,\mathsf{b}^0_i)_{\mathbb{R}^{p-|I_0|}}=
\lim\limits_{j\to\infty}
(\mathsf{b}^{\e_j}_k,\mathsf{b}^{\e_j}_i)_{\mathbb{R}^{p-|I_0|}}
=\d_{ki},
\end{equation*}
where $\d_{ki}$ is the Kronecker delta. Hence, $p$ vectors
$\mathsf{b}_k^0$ of dimensions $(p-|I_0|)<p$ is an
orthonormalized system, a contradiction. The proof of
Theorem~\ref{th1.2} is complete.

Let us clarify the asymptotics behaviour of the perturbed
eigenfunctions under hypothesis of Theorem~\ref{th1.2}. Let
$\l_0=\l_0^k$ be a simple limiting eigenvalue. It follows from
(\ref{3.34}) and Lemmas~\ref{lm3.5},~\ref{lm3.6} that
\begin{equation*}
\mathsf{b}_{kk}^\e=(\psi_0^k+\e\phi_1^k\cos
Mx_3,\psi_\e^k)_{L_2(\Om)}+O(\e^{3/2}(|\ln\eta|^{3/2}+1)).
\end{equation*}
Therefore, due to (\ref{3.31}), (\ref{3.33}) and
Lemmas~\ref{lm3.5},~\ref{lm3.6} the eigenfunction of the
perturbed problem
\begin{equation}\label{3.35}
\t\psi_\e^k=(\psi_0^k+\e\phi_1^k\cos
Mx_3,\psi_\e^k)_{L_2(\Om)}\psi_\e^k,
\end{equation}
associated with $\l_\e^k$, has the asymptotics
\begin{equation}\label{3.36}
\begin{aligned}
\t\psi_\e^k(x)=\psi_0^k(x)&+\e\phi_1^k(x',\eta,\e)\cos Mx_3 +
\\
{}&+\e\chi(\tau/c_0)\psi_0^{k,\nu}(s)X(\xi,\eta
\mathsf{g}_\e(s)) +O\left(\e^{3/2}(|\ln\eta|^{3/2}+1)\right),
\end{aligned}
\end{equation}
in $H^1(\Om)$, where $\phi_0^{k,\nu}$ is a value of normal
derivative of the function $\phi_0^k$ on the boundary $\p\om$.
The asymptotics obtained and Lemmas~\ref{lm3.4}--\ref{lm3.6}
yield that $\|\t\psi_\e^k-\psi_0^k\|_{H^1(\Om)}=o(1)$.

Let $\l_0=\l_0^q=\ldots=\l_0^{q+p-1}$ be a $p$-multiply limiting
eigenvalue. Like before, by (\ref{3.34}) and
Lemmas~\ref{lm3.5},~\ref{lm3.6} we deduce:
\begin{equation*}
\mathsf{b}_{ki}^\e=(\psi_0^k+\e\phi_1^k\cos
Mx_3,\psi_\e^i)_{L_2(\Om)}+O(\e^{3/2}(|\ln\eta|^{3/2}+1)).
\end{equation*}
From this fact, (\ref{3.31}), (\ref{3.33}) and
Lemmas~\ref{lm3.5},~\ref{lm3.6} it follows that a linear
combination of the perturbed eigenfunctions
\begin{equation}\label{3.37}
\t\psi_\e^k=\sum\limits_{i=q}^{q+p-1}(\psi_0^k+\e\phi_1^k\cos
Mx_3,\psi_\e^i)_{L_2(\Om)}\psi_\e^i
\end{equation}
obeys  asymptotics (\ref{3.36}) in a sense of  $H^1(\Om)$ norm,
where by $\t\psi_\e^k$ we mean the function (\ref{3.37}). In
particular, this fact implies that the functions $\t\psi_\e^k$
from  (\ref{3.37}) satisfies an equality
$\|\t\psi_\e^k-\psi_0^k\|_{H^1(\Om)}=o(1)$. Thus, we have proved
\begin{theorem}\label{th3.1}
Suppose the hypothesis of Theorem~\ref{th1.2} holds. Then for
each eigenfunction $\psi_0^k$ of the limiting problem there
exists a perturbed eigenfunction $\t\psi_\e^k$ from (\ref{3.35})
if limiting eigenvalue $\l_0^k$ is a simple and a linear
combination $\t\psi_\e^k$ from (\ref{3.37}) composed by
eigenfunctions $\psi_\e^i$, $i=q,\ldots,q+p-1$ if limiting
eigenvalue $\l_0=\l_0^q=\ldots=\l_0^{q+p-1}$ is $p$-multiply,
and this function or combination satisfies the equality
$\|\t\psi_\e^k-\psi_0^k\|_{H^1(\Om)}=o(1)$ and has the
asymptotics (\ref{3.36}) in $H^1(\Om)$ norm.
\end{theorem}

\sect{Proof of Theorems~\ref{th1.3},~\ref{th1.4}}

\PF{\indent Proof of Theorem~\ref{th1.3}.} Throughout the proof,
if it is not said specially, we keep the notations from the
previous section. Since $\l_0^k$ is a double  eigenvalue, after
the arranging (\ref{1.10}) it will appear twice in the sequence
$\{\l_0^j\}_{j=1}^\infty$; assume that $\l_0^k=\l_0^{k+1}$. Then
$\k\equiv\k_k=\k_{k+1}$, $M_k=M_{k+1}$. Associated
eigenfunctions counting all normalization and orthogonalization
prescribed in the first section read as follows:
\begin{align*}
{}&\phi_0^k(x')=\frac{2}{\pi(\mathcal{J}'_n(\sqrt{\k}))^2}
\mathcal{J}_n(\sqrt{\k}r)\cos(n\th+\a_\e),
\\
{}&\phi_0^{k+1}(x')=\frac{2}{\pi(\mathcal{J}'_n(\sqrt{\k}))^2}
\mathcal{J}_n(\sqrt{\k}r)\sin(n\th+\a_\e),
\\
{}&\psi_0^k(x)=\phi_0^k(x')\cos Mx_3,\quad
\psi_0^{k+1}(x)=\phi_0^{k+1}(x')\cos Mx_3.
\end{align*}
The equation (\ref{1.12}) for $\a_\e$, as one can easily check,
is solvable and it is exactly the condition of orthogonality for
normal derivatives of the functions $\phi_0^k$ and
$\phi_0^{k+1}$ in $L_2(\p\om)$ weighted by
$(-\ln\sin\eta\mathsf{g}_\e)$. Leading terms  of the asymptotics
for the eigenvalues $\l_\e^k$ and $\l_\e^{k+1}$, in accordance
with Theorem~\ref{th1.2}, are
\begin{align*}
{}&\l_1^k(\eta,\e)=\frac{2\k}{\pi}\int\limits_0^{2\pi}
\sin^2(n\th+\a_\e)\ln\sin\eta \mathsf{g}_\e(\th)
\,\mathrm{d}\th,
\\
{}&\l_1^{k+1}(\eta,\e)=
\frac{2\k}{\pi}\int\limits_0^{2\pi}\cos^2(n\th+\a_\e)
\ln\sin\eta \mathsf{g}_\e(\th)\,\mathrm{d}\th.
\end{align*}
Let us prove that these terms are same:
\begin{align*}
\l_1^k-\l_1^{k+1}&=\frac{2\k}{\pi}\int\limits_{\p\om}
\cos(2n\th+2\a_\e)\ln\sin\eta \mathsf{g}_\e(\th)\,\mathrm{d}\th=
\\
{}&=\frac{2\k}{\pi}\int\limits_0^{2\pi}\sin(2nt+\a_\e)
\ln\sin\eta \mathsf{g}_\e(t)\,\mathrm{d}t=0.
\end{align*}
In calculations we had made a change $t=\th-\pi/(2n)$, after
that we used $\pi/(2n)$-periodicity of $\mathsf{g}_\e$ and the
equality (\ref{1.12}). Now we are going to prove that
$\l_\e^k=\l_\e^{k+1}$. Suppose it is wrong, then $\l_\e^k$,
$\l_\e^{k+1}$ are simple eigenvalues. According with
Theorem~\ref{th3.1}, for the functions $\psi_0^k$,
$\psi_0^{k+1}$ there exist linear combination of the
eigenfunctions $\psi_\e^k$, $\psi_\e^{k+1}$, converging to
$\psi_0^k$, $\psi_0^{k+1}$ in $H^1(\Om)$:
\begin{equation}\label{4.1}
\mathsf{c}_1^\e\psi_\e^k+\mathsf{c}_2^\e\psi_\e^{k+1}\to\psi_0^k,
\quad
\mathsf{c}_3^\e\psi_\e^k+\mathsf{c}_4^\e\psi_\e^{k+1}\to\psi_0^{k+1}.
\end{equation}
From the hypothesis of the theorem it follows that
$\psi_\e^k(r,\th+\pi/(2n),x_3)$ and
$\psi_\e^{k+1}(r,\th+\pi/(2n),x_3)$ are perturbed eigenfunctions
associated with $\l_\e^k$ and $\l_\e^{k+1}$, therefore
\begin{equation*}
\psi_\e^k(r,\th+\pi/(2n),x_3)=\mathsf{c}^\e_5
\psi_\e^k(r,\th,x_3),\quad \psi_\e^{k+1}(r,\th+\pi/(2n),x_3)=
\mathsf{c}^\e_6 \psi_\e^{k+1}(r,\th,x_3).
\end{equation*}
The equalities obtained and (\ref{4.1}) yield
\begin{equation}\label{4.2}
\mathsf{c}_1^\e\mathsf{c}_5^\e\psi_\e^k+
\mathsf{c}_2^\e\mathsf{c}_6^\e\psi_\e^{k+1}\to-\psi_0^{k+1},
\quad \mathsf{c}_3^\e\mathsf{c}_1^\e\psi_\e^k+
\mathsf{c}_4^\e\mathsf{c}_6^\e\psi_\e^{k+1}\to\psi_0^{k}.
\end{equation}
Now we multiply first convergence from (\ref{4.1}) by the second
from (\ref{4.2}) in  $L_2(\Om)$ and we do the same with second
convergence from  (\ref{4.1}) and the first from (\ref{4.2}).
The result reads as follows:
\begin{equation*}
\mathsf{c}_1^\e\mathsf{c}_3^\e\mathsf{c}_5^\e+
\mathsf{c}_2^\e\mathsf{c}_4^\e\mathsf{c}_6^\e\psi_\e^{k+1}\to
H/2,\quad \mathsf{c}_1^\e\mathsf{c}_3^\e\mathsf{c}_5^\e+
\mathsf{c}_2^\e\mathsf{c}_4^\e\mathsf{c}_6^\e\psi_\e^{k+1}\to
-H/2,
\end{equation*}
a contradiction, i.e., $\l_\e=\l_\e^{k}=\l_\e^{k+1}$ is a double
eigenvalue. The asymptotics of associated eigenfunctions can be
easily obtained from Theorem~\ref{th3.1}; linear combinations
converging to $\psi_0^k$ and $\psi_0^{k+1}$, owing to $\l_\e$
being double are associated eigenfunctions. The main terms of
the asymptotics from Theorem~\ref{th3.1} depends on $\e$, what
happens because of additional orthogonalization in $L_2(\p\om)$.
At the same time, it is easy to eliminate this dependence: we
just should consider suitable linear combinations of the
functions $\t\psi_\e^k$ and $\t\psi_\e^{k+1}$ from
Theorem~\ref{th3.1}; their main terms should be
$\mathcal{J}_n(\sqrt{\k}r)\cos n\th$ and
$\mathcal{J}_n(\sqrt{\k}r)\sin n\th$. As a result of these
simple calculations we conclude that eigenfunctions associated
with $\l_\e$ can be chosen such that they converge to
$\mathcal{J}_n(\sqrt{\k}r)\cos n\th\cos Mx_3$ and
$\mathcal{J}_n(\sqrt{\k}r)\sin n\th\cos Mx_3$ in $H^1(\Om)$ and
have in  $H^1(\Om)$-norm asymptotics
\begin{equation}\label{4.3}
\begin{aligned}
\t\psi_\e^k(x&)=\cos Mx_3\Big(\mathcal{J}_n(\sqrt{\k}r)\cos
n\th+\e\t\psi_1^k(x',\eta,\e)+
\\
{}&+\e\chi(\tau/c_0)\sqrt{\k}\mathcal{J}'_n(\sqrt{\k})
X(\xi,\eta \mathsf{g}_\e(s))\cos
n\th\Big)+O(\e^{3/2}(|\ln\eta|^{3/2}+1)),
\\
\t\psi_\e^{k+1}(x&)=\cos Mx_3\Big(\mathcal{J}_n(\sqrt{\k}r)\sin
n\th+\e\t\psi_1^{k+1}(x',\eta,\e)+
\\
{}&+\e\sqrt{\k}\mathcal{J}'_n(\sqrt{\k})\chi(\tau/c_0)X(\xi,\eta
\mathsf{g}_\e(s))\sin n\th\Big)+O(\e^{3/2}(|\ln\eta|^{3/2}+1)),
\end{aligned}
\end{equation}
where $\t\phi_\e^k$ and $\t\phi_\e^{k+1}$ are solutions to the
problem (\ref{3.4}), (\ref{3.12}) with $\l_1=\l_1^k=\l_1^{k+1}$,
$\phi_0(x')=\mathcal{J}_n(\sqrt{\k}r)\cos n\th$ and
$\phi_0(x')=\mathcal{J}_n(\sqrt{\k}r)\sin n\th$, respectively.
The proof is complete.

\smallskip

\PF{\indent Proof of Theorem~\ref{th1.4}.} Let
\begin{equation*}
\g_{\e,*}=\left\{x: x'\in\p\om, \left|x_3-
\e\pi(j+1/2)\right|<\e c\eta, j=0,\ldots,N-1\right\},
\end{equation*}
$\l_{\e,*}^k$ indicate eigenvalues of the problem (\ref{1.2}),
(\ref{1.3}) with $\g_\e$ and $\G_\e$ replaced by  $\g_{\e,*}$
and $\Si\backslash\overline{\g}_{\e,*}$, respectively. The set
$\g_{\e,*}$ satisfies the hypothesis of Theorem~\ref{th1.2} with
the functions $\mathsf{g}_\e\equiv1$, hence, $\l_{\e,*}^k$ meet
asymptotics (\ref{1.5}). The definition of $\g_\e$ implies that
$\g_{\e,*}\subseteq\g_\e$; it is also clear that
$\g_\e\subseteq\Si$. Using these inclusions and minimax
properties of eigenvalues for elliptic problem it is easy to
show that
\begin{equation*}
\l_{\e,*}^k\le\l_\e^k\le\l_0^k,
\end{equation*}
from what, the asymptotics (\ref{1.5}) for $\l_{\e,*}^k$ and the
inequalities (\ref{3.27}) we get the maintained estimates for
degree of convergence. The proof is complete.


\renewcommand{\refname}

\end{document}